# PyAutoLens: Open-Source Strong Gravitational Lensing

**James. W. Nightingale**[1], **Richard G. Hayes**[1], **Ashley Kelly**[1], **Aristeidis Amvrosiadis**[1], **Amy Etherington**[1], **Qiuhan He**[1], **Nan Li**[2], **XiaoYue Cao**[3], **Jonathan Frawley**[4], **Shaun Cole**[1], **Andrea Enia**[5], **Carlos S. Frenk**[1], **David R. Harvey**[6], **Ran Li**[3], **Richard J. Massey**[1], **Mattia Negrello**[7], and **Andrew Robertson**[1]

**1** Institute for Computational Cosmology, Stockton Rd, Durham DH1 3LE **2** Key Laboratory of Space Astronomy and Technology, National Astronomical Observatories, Chinese Academy of Sciences, Beijing 100101, China **3** National Astronomical Observatories, Chinese Academy of Sciences, 20A Datun Road, Chaoyang District, Beijing 100012, China **4** Advanced Research Computing, Durham University, Durham DH1 3LE **5** Dipartimento di Fisica e Astronomia, Università degli Studi di Bologna, Via Berti Pichat 6/2, I-40127 Bologna, Italy **6** Lorentz Institute, Leiden University, Niels Bohrweg 2, Leiden, NL-2333 CA, The Netherlands **7** School of Physics and Astronomy, Cardiff University, The Parade, Cardiff CF24 3AA, UK

## Summary

Strong gravitational lensing, which can make a background source galaxy appears multiple times due to its light rays being deflected by the mass of one or more foreground lens galaxies, provides astronomers with a powerful tool to study dark matter, cosmology and the most distant Universe. `PyAutoLens` is an open-source Python 3.6+ package for strong gravitational lensing, with core features including fully automated strong lens modeling of galaxies and galaxy clusters, support for direct imaging and interferometer datasets and comprehensive tools for simulating samples of strong lenses. The API allows users to perform ray-tracing by using analytic light and mass profiles to build strong lens systems. Accompanying `PyAutoLens` is the autolens workspace, which includes example scripts, lens datasets and the `HowToLens` lectures in Jupyter notebook format which introduce non-experts to strong lensing using `PyAutoLens`. Readers can try `PyAutoLens` right now by going to the introduction Jupyter notebook on Binder or checkout the readthedocs for a complete overview of `PyAutoLens`'s features.

## Background

When two galaxies are aligned down the line-of-sight to Earth, light rays from the background galaxy are deflected by the intervening mass of one or more foreground galaxies. Sometimes its light is fully bent around the foreground galaxies, traversing multiple paths to the Earth, meaning that the background galaxy is observed multiple times. This alignment of galaxies is called a strong gravitational lens, an example of which, SLACS1430+4105, is shown in the image below. The massive elliptical lens galaxy can be seen in the centre of the left panel, surrounded by a multiply imaged source galaxy whose light has been distorted into an Einstein ring. The central panel shows a `PyAutoLens` reconstruction of the lensed source's light, where the foreground lens's light was simultaneously fitted for and subtracted to reveal the source. The right panel shows a pixelized reconstruction of the source's unlensed light distribution performed by `PyAutoLens`, which is created using a model of the lens galaxy's mass to trace backwards how the source's light is gravitationally lensed.

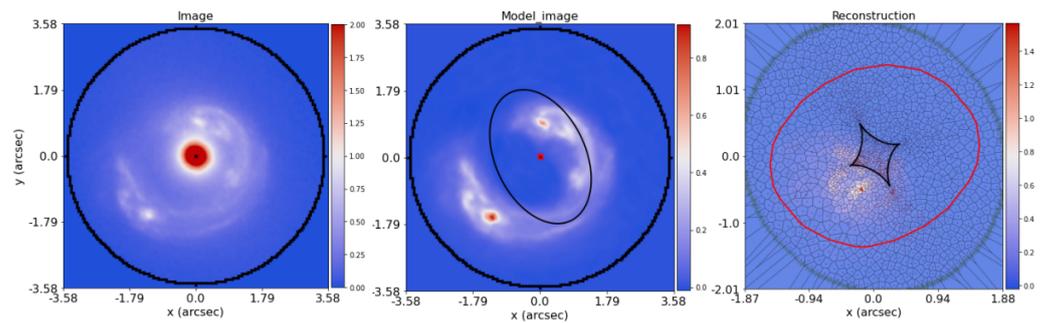

**Figure 1:** Hubble Space Telescope imaging of the strong lens SLACSJ1430+1405 (left column), a fit to its lensed source galaxy (middle column) and unlensed source reconstruction (right column) using `PyAutoLens`.

Strong lensing provides astronomers with an invaluable tool to study a diverse range of topics. Mass modeling of strong lenses has quantified the distribution of stars (Koopmans et al., 2009) (Sonnenfeld et al., 2015) (Treu et al., 2009) (Nightingale et al., 2019) and invisible dark matter (Vegetti et al., 2014) of galaxies. The source galaxy is highly magnified and reconstruction of its light allows a view of fainter or more distant objects than would otherwise be possible (Dye et al., 2014) (Enia et al., 2018). Strong lensing is a competitive test of cosmological models, for example the expansion rate of the Universe can be inferred from the 'time-delay' between different image paths to the same distant quasar (Suyu et al., 2017). Strong lensing of galaxy clusters has also made many contributions to all these topics (Jullo et al., 2010) (Richard et al., 2014) (Atek et al., 2015).

## Statement of Need

The past decade has seen the discovery of many hundreds of galaxy-scale and cluster-scale lenses, with high quality imaging (Bolton et al., 2012), interferometer (Negrello et al., 2014) (Enia et al., 2018) and spectroscopy (Czoske et al., 2012) datasets now available. Historically, the modeling of a strong lens is a time-intensive process that requires significant human intervention to perform, restricting the scope and size of the scientific analysis. In the next decade of order of *one hundred thousand* strong lenses will be discovered by surveys such as Euclid, LSST and SKA (Collett, 2015), demanding a widely available and automated approach for strong lens analysis. `PyAutoLens` aims to meet this need, by making strong lens analysis accessible to the wider Astronomy community and enabling the automated analysis of large samples of strong lenses.

## Software API and Features

A gravitational lens system can be quickly assembled from Python objects which provide abstract data representations of the different components of a strong lens. A `Galaxy` object contains one or more `LightProfile`'s and `MassProfile`'s, which represent its two dimensional distribution of starlight and mass. `Galaxy`'s lie at a particular distance (redshift) from the observer, and are grouped into `Plane`'s. Raytracing through multiple `Plane`s is achieved by passing them to a `Tracer` with an `astropy` Cosmology. By passing any of these objects a `Grid2D` object strong lens quantities can be computed, including multi-plane ray-tracing sightlines (McCully et al., 2014). All of these objects are extensible, making it straightforward to compose highly customized lensing system. Ray-tracing calculations are optimized using the packages NumPy (van der Walt et al., 2011), numba (Lam et al., 2015) and pyquad (Kelly, 2020).

To perform lens modeling, PyAutoLens adopts the probabilistic programming language PyAutoFit (https://github.com/rhayes777/PyAutoFit). PyAutoFit allows users to compose a lens model from LightProfile, MassProfile and Galaxy objects, customize the model parameterization and fit it to data via a NonLinearSearch (e.g., dynesty (Speagle, 2020), emcee (Foreman-Mackey et al., 2013), PySwarms (Miranda, 2018)). By composing a lens model with a Pixelization and Regularization object, the background source's light is reconstructed using a rectangular grid or Voronoi mesh that accounts for irregular galaxy morphologies. Lensed quasar and supernovae datasets can be fitted using a PointSource, which uses their observed positions, flux-ratios and time-delays to fit the lens model. Strong lensing clusters consisting of any number of lens galaxies can also be analysed with PyAutoLens using these objects.

Automated lens modeling uses PyAutoFit's non-linear search chaining feature, which breaks the model-fit into a chained sequence of non-linear searches. These fits pass information gained about simpler lens models fitted by earlier searches to subsequent searches, which fit progressively more complex models. By granularizing the model-fitting procedure, automated pipelines that fit complex lens models without human intervention can be carefully crafted, with example pipelines found on the autolens workspace. To ensure the analysis and interpretation of fits to large lens datasets is feasible, PyAutoFit's database tools write lens modeling results to a relational database which can be loaded from hard-disk to a Python script or Jupyter notebook. This uses memory-light Python generators, ensuring it is practical for thousands of lenses.

PyAutoLens includes a comprehensive visualization library for the analysis of both direct imaging and submm / radio interferometer datasets, tools for preprocessing data to formats suited to lens analysis and options to include effects like the telescope optics and background sky subtraction in the model-fit. Interferometer analysis is performed directly on the observed visibilities in their native Fourier space, circumventing issues associated with the incomplete sampling of the uv-plane that give rise to artefacts that can bias the inferred mass model and source reconstruction in real-space. To make feasible the analysis of millions of visibilities, PyAutoLens uses PyNUFFT (Lin, 2018) to fit the visibilities via a non-uniform fast Fourier transform and PyLops (Ravasi & Vasconcelos, 2019) to express the memory-intensive linear algebra calculations as efficient linear operators (Powell et al., 2020). Creating realistic simulations of imaging and interferometer strong lensing datasets is possible, as performed by (Alexander et al., 2019) (Hermans et al., 2019) who used PyAutoLens to train neural networks to detect strong lenses.

## Performance

The analysis of direct imaging datasets and interferometer datasets (up to of order 1 million visibilities) are both feasible on hardware with at least 4GB of RAM. The time it takes to perform lens modeling with PyAutoLens is highly variable and depends on the size of the dataset being analysed and complexity of the model being fitted. They can vary from minutes to thousands of CPU hours. The run-times section on readthedocs provides graphs showing the performance of the latest release of PyAutoLens and a calculator for estimating how long a lens model fit may take. For large jobs we recommend users install PyAutoLens on a HPC cluster and documentation is provided on how to set this up.

## Workspace and HowToLens Tutorials

PyAutoLens is distributed with the autolens workspace, which contains example scripts for modeling and simulating strong lenses and tutorials on how to preprocess imaging and interferometer datasets before a PyAutoLens analysis. Also included are the HowToLens tutorials, a five chapter lecture series composed of over 30 Jupyter notebooks aimed at non-experts, introducing them to strong gravitational lensing, Bayesian inference and teaching them how to use `PyAutoLens` for their scientific study. The lectures are available on our Binder and may therefore be taken without a local `PyAutoLens` installation.

## Software Citations

`PyAutoLens` is written in Python 3.6+ (Van Rossum & Drake, 2009) and uses the following software packages:

- `Astropy` (Astropy Collaboration et al., 2013) (Price-Whelan et al., 2018)
- `COLOSSUS` (Diemer, 2018)
- `corner.py` (Foreman-Mackey, 2016)
- `dynesty` (Speagle, 2020)
- `emcee` (Foreman-Mackey et al., 2013)
- `Matplotlib` (Hunter, 2007)
- `numba` (Lam et al., 2015)
- `NumPy` (van der Walt et al., 2011)
- `PyAutoFit` (Nightingale et al., 2021)
- `PyLops` (Ravasi & Vasconcelos, 2019)
- `PyMultiNest` (Buchner et al., 2014) (Feroz et al., 2009)
- `PyNUFFT` (Lin, 2018)
- `pyprojroot` (https://github.com/chendaniely/pyprojroot)
- `pyquad` (Kelly, 2020)
- `PySwarms` (Miranda, 2018)
- `scikit-image` (Van der Walt et al., 2014)
- `scikit-learn` (Pedregosa et al., 2011)
- `Scipy` (Virtanen et al., 2020)

## Related Software

- `AutoLens` (Nightingale & Dye, 2015) (Nightingale et al., 2018)
- `gravlens` http://www.physics.rutgers.edu/~keeton/gravlens/manual.pdf
- `lenstronomy` https://github.com/sibirrer/lenstronomy (Birrer & Amara, 2018)
- `visilens` https://github.com/jspilker/visilens (Spilker et al., 2016)

## Acknowledgements

JWN and RJM are supported by the UK Space Agency, through grant ST/V001582/1, and by InnovateUK through grant TS/V002856/1. RGH is supported by STFC Opportunities grant ST/T002565/1. QH, CSF and SMC are supported by ERC Advanced In-vestigator grant, DMIDAS [GA 786910] and also by the STFCConsolidated Grant for Astronomy at Durham [grant numbersST/F001166/1, ST/I00162X/1,ST/P000541/1]. RJM is supported by a Royal Society University Research Fellowship. DH acknowledges support by the ITP Delta foundation. AR is supported bythe ERC Horizon2020 project 'EWC' (award AMD-776247-6). MN has received funding from the European Union's Horizon 2020 research and innovation programme under the Marie Sklodowska-Curie grant agreement no. 707601. This work used the DiRAC@Durham facility managed by the Institute for Computational Cosmology on behalf of the STFC DiRAC HPC Facility (www.dirac.ac.uk). The equipment was funded by BEIS capital funding via STFC capital grants ST/K00042X/1, ST/P002293/1, ST/R002371/1 and ST/S002502/1, Durham University and STFC operations grant ST/R000832/1. DiRAC is part of the National e-Infrastructure.